\documentstyle[12pt,epsf,procsla]{article}
\begin{document}

\centerline{\normalsize\bf MONOPOLE CONDENSATION IN}
\baselineskip=22pt
\centerline{\normalsize\bf GAUGE THEORY VACUUM}
\baselineskip=16pt

\centerline{\footnotesize ADRIANO DI GIACOMO}
\baselineskip=13pt
\centerline{\footnotesize\it University Pisa \& I.N.F.N. ITALY}
\baselineskip=12pt
\centerline{\footnotesize\it Pisa, 56100, Italy}
\centerline{\footnotesize E-mail: digiaco@ipifidpt.difi.unipi.it}

\vspace*{0.9cm}
\abstracts{The condensation of monopoles
(dual superconductivity)
of QCD vacuum is reviewed.
Direct evidence is produced that the system, in the confined phase,
is a dual superconductor.}

\normalsize\baselineskip=15pt
\setcounter{footnote}{0}
\renewcommand{\thefootnote}{\alph{footnote}}

\section{Introduction: statement of the problem}

Dual superconductivity of the vacuum was advocated long ago as the mechanism
for
confinement of colour\cite{1,2,3}.
Dual means that the role of electric and magnetic fields and charges are
interchanged with respect to ordinary superconductors. The basic idea is that
the
chromoelectric field acting between a quark antiquark pair is channeled into an
Abrikosov flux tube\cite{4}, by dual Meissner effect. The resulting static
potential is proportional to the distance $R$
\begin{equation}
V(R) = \sigma R \label{eq1.1}\end{equation}
$\sigma$ is the string tension. Flux tubes are expected to behave as
strings\cite{5,6}.

Numerical simulations of QCD on the lattice support this picture:
\begin{itemize}
\item[1)] The interquark force at large distances obeys
Eq.(\ref{eq1.1})\cite{7}.
\item[2)] Flux tubes exist in field configurations produced by static $q\bar q$
pairs\cite{8,9,10}.
\item[3)] Higher modes of the string are visible\cite{11}.
\end{itemize}
Till recently, however, a convincing demonstration that the ground state of QCD
behaves as a superconductor was still lacking. In the following I will analyse
recent progress on
this point. In particular I will present direct
evidence of dual superconductivity of QCD vacuum, obtained by measuring
on a  lattice a  disorder parameter\cite{12}.

Ordinary superconductivity is nothing but the spontaneous breaking (S.B.), \`a
la
Higgs, of the $U(1)$ symmetry related to charge conservation\cite{13}. A
charged
field
\begin{equation}
\Phi = \psi\,{\rm e}^{{\rm i} \theta q}\qquad \psi =
|\Phi|\label{eq1.2}\end{equation}
acquires a non vanishing vacuum expectation value (v.e.v.)
$\langle\Phi\rangle$.
As a consequence
\begin{itemize}
\item[(i)] the photon acquires a mass $\mu$
\begin{equation}
\mu^2 = e^2\,\langle\Phi\rangle^2\label{eq1.3}\end{equation}
\item[(ii)] the vacuum
is not $U(1)$ invariant, and has no definite charge: indeed
if
it where invariant the v.e.v. of any charged operator would vanish.
\end{itemize}
A well known consequence of the Higgs phenomenon is that the derivative of the
angular variable $\theta$ of Eq.(\ref{eq1.2}) becomes the longitudinal
component of
the photon. Instead of $A_\mu$ it proves convenient to use as a field variable
$\tilde A_\mu = A_\mu - \frac{1}{e}\partial_\mu\theta$ which is gauge
invariant.
In terms of $\tilde A_\mu$ $F_{\mu\nu} = \partial_\mu\tilde A_\nu -
\partial_\nu\tilde A_\mu$: in particular ${\bf H} = {\bf \nabla}\wedge{\bf
\tilde
A}$. The equations of motion for a static configuration become
\begin{equation}
\partial_i F_{ij} + \mu^2 \tilde A_j = 0 \label{eq1.4}\end{equation}
or
\begin{equation}
{\bf \nabla}\wedge{\bf H} = \mu^2 {\bf \tilde A} \label{eq1.5}\end{equation}
Taking the curl of both sides of Eq.(\ref{eq1.5}) gives
\begin{equation}
{\bf \nabla}^2 {\bf H} + \mu^2 {\bf H} = 0 \label{eq1.6}\end{equation}
Eq.(\ref{eq1.5}) means that a permanent current (London current)
\begin{equation}
{\bf j} = \mu^2 {\bf \tilde A}\label{eq1.7}\end{equation}
is present in the superconductor, with ${\bf E} = 0$, or, since $\rho {\bf j} =
{\bf E}$, that $\rho = 0$.

Eq.(\ref{eq1.6}) means that the magnetic field ${\bf H}$ has a finite
penetration
depth, and this is nothing but Meissner effect. On a line around a flux tube at
distance larger than the penetration depth ${\bf \tilde A} = 0$,
$\oint {\bf \tilde A} d{\bf x} = 0$ or, by the definition of ${\tilde A}_\mu$
$\oint {\bf A} d{\bf x} = n\pi/q$, which is flux quantization. The key
parameter in
the game is $\langle \Phi\rangle$. To detect superconductivity one can either
look
for permanent currents Eq.(\ref{eq1.7}), i.e. demonstrate that $\mu^2\neq 0$,
or
directly for the spontaneous breaking of $U(1)$, i.e.
look
for a non vanishing  v.e.v. of a charged operator.

In QCD the dual situation is expected to occur. The disorder parameter is the
v.e.v. of an operator with non zero magnetic charge, and the London current is
a magnetic current.
The strategy of detecting
dual superconductivity by looking for
persistent currents will be reviewed by D. Haymaker in
his talk to this conference. I will instead present a direct
determination of the disorder parameter $\langle \Phi\rangle$.

\section{Monopoles in gauge theories}
Monopoles as solitons in gauge theories are related to the elements of the
first
homotopy group of the gauge group\cite{14}. Since $\Pi_1(SU(N)) = \{1\}$ in
order
to have monopoles the symmetry has to reduce to some non simply connected
group.
In a theory with $SU(2)$ gauge group coupled to a scalar field  in the
adjoint representation\cite{15} $\vec \Phi$, when the Higgs phenomenon reduces
the
symmetry from $SU(2)$ to $U(1)$  monopoles do exist as stable static
solutions\cite{16,17}. The relevant degrees of freedom are described by the
gauge
invariant field strength\cite{16}
\begin{equation}
f_{\mu\nu} = {\vec G}_{\mu\nu}\cdot\vec \Phi - \frac{1}{g}
\hat\Phi\cdot\left( D_\mu\hat \Phi\wedge D_\nu\hat\Phi\right)
\label{eq2.1}\end{equation}
$\hat\Phi = \vec \Phi/|\vec \Phi|$ is the colour direction of the Higgs field.
At large distances the field $f_{\mu\nu}$ of a monopole configuration is the
field
of a Dirac monopole of magnetic charge 2. One can define a gauge field $a_\mu$
\begin{equation}
a_\mu = {\vec A}_\mu\cdot\hat \Phi
\label{eq2.2}\end{equation}
Contrary to $f_{\mu\nu}$ $a_\mu$ is not gauge invariant, since ${\vec A}_\mu$
is
not gauge covariant. In general\cite{18}
\begin{equation}
f_{\mu\nu} = \partial_\mu a_\nu - \partial_\nu a_\mu -
\frac{1}{g}\hat \Phi\left(\partial_\mu\hat\Phi\wedge \partial_\nu\hat
\Phi\right)
\label{eq2.3}\end{equation}
After a gauge rotation which brings $\hat \Phi$ in a given colour direction
$(\hat\Phi)^a = \delta^a_3$, the last term in Eq.(\ref{eq2.3}) vanishes and
\begin{equation}
f_{\mu\nu} = \partial_\mu a_\nu - \partial_\nu a_\mu
\label{eq2.4}\end{equation}
Such a gauge rotation is called an abelian projection: in a gauge defined by
this
procedure the $U(1)$ degrees of freedom relevant to the definition of monopoles
coincide with a subgroup of the gauge group. $a_\nu$ and $f_{\mu\nu}$ are
formally
identical to the fields of a $U(1)$ gauge theory. We notice for further
reference
that also the commutation relations between $f_{0i}$ and $a_i$ are identical to
those of a $U(1)$ theory.

To define the monopoles which produce, by condensation in the
vacuum, dual superconductivity and confinement, the relevant degrees of freedom
have
to be selected by an abelian projection\cite{19}. A few different abelian
projections have been proposed in the literature as candidates for this
purpose\cite{19,20} and will be discussed in detail in what follows.

We conclude this section by noticing that, whatever the relevant abelian
projection, the problem is always reduced to detect dual superconductivity of a
$U(1)$ system.

\section{Detecting dual superconductivity in $U(1)$ gauge theory}
I will sketch the construction of the creation operator for a
monopole\cite{12},
whose v.e.v. will be used as a disorder parameter for dual superconductivity.
Let
$\Pi_i({\bf x},t) = F_{0i}({\bf x},t)$ be the usual conjugate momenta to the
field variables $A_i({\bf x},t)$. The operator
\begin{equation}
\mu({\bf y},t) = {\rm exp}\left({\rm i}\int d^3{\bf x} {\bf \Pi}({\bf x},t)
\frac{1}{e}{\bf b}({\bf x}-{\bf y})\right)
\label{eq3.1}\end{equation}
creates a monopole of magnetic charge $m$ in the site ${\bf y}$ at time $t$, if
$\frac{1}{e}{\bf b}({\bf x}-{\bf y})$ is the classical vector potential
produced
by such a monopole, with the Dirac string subtracted. Putting the string along
the
direction ${\bf n}$
\begin{equation}
{\bf b}({\bf r}) = \frac{m}{2}\frac{\displaystyle {\bf n}\wedge{\bf r}}
{\displaystyle r(r - {\bf n}{\bf r})}
\label{eq3.2}\end{equation}
Indeed $\mu$ as defined by Eq.(\ref{eq3.1}) is the operator which adds to any
field configuration the field of the monopole, in the same ad the translation
operator  adds $a$ to the position $q$:
\[ {\rm e}^{i p a} | q\rangle = |q + a\rangle\]
$\mu({\bf y},t)$ carries magnetic charge $m$. By use of the canonical
commutation
relation $[ \Pi_i({\bf x},t, A_j({\bf y},t)] = -{\rm i}\delta_{ij}\delta^3(
{\bf x} - {\bf y})$, and of the definition of the magnetic charge operator
\begin{equation}
Q = \int d^3{\bf x} {\bf \nabla}\cdot{\bf H} =
\int d^3{\bf x} {\bf \nabla}\cdot ({\bf \nabla}\wedge {\bf
A})\label{eq3.3}\end{equation}
\begin{equation}
\left[ Q,\mu({\bf y},t)\right] =
\int d^3{\bf x} \frac{1}{e}{\bf \nabla}\cdot ({\bf \nabla}\wedge {\bf b})
\mu({\bf y},t) = \frac{2\pi m}{e} \mu({\bf y},t)
\label{eq3.4}\end{equation}
We will use the v.e.v. $\langle\mu\rangle$ as disordere parameter for dual
superconductivity\cite{12}. Our construction is inspired by the classical work
of
ref.\cite{21} and by its application to monopole condensation of ref.\cite{22}.
In
ref.\cite{22} condensation of monopoles is proved, in the infinite volume
limit,
for a specific form of the action, the Villain action. Our construction
coincides
with ref.\cite{22} for that case, but can be used for any form of the action,
and
for finite volumes. The infinite volume limit can be reached by finite size
analysis. I refer to ref.\cite{12} for the details of the construction which I
will summarize as follows.

\begin{itemize}
\item[i)] $\langle\mu\rangle$ can be determined either by the cluster property
from the correlation of a monopole and an antimonopole at large distance $d$
\begin{equation}
\langle \mu({\bf d},0)\,\bar\mu({\bf 0},0)\rangle \mathop\simeq\limits_{
|{\bf d}|\to \infty} \langle \mu\rangle ^2
\label{eq3.5}\end{equation}
or directly. It is known that, for Wilson action on lattice, electric charge is
confined for $\beta < \beta_c$ ( $\beta= 1/e^2$, $\beta_c \simeq 1.01$); for
$\beta > \beta_c$ the system is made of free photons. We expect
$\langle \mu\rangle_{V\to \infty}\neq 0$ for $\beta < \beta_c$ and
$\langle \mu\rangle_{V\to \infty} = 0$ for $\beta > \beta_c$.
Of course $\langle \mu\rangle$ beeing an analytic function of $\beta$ for
finite
volume, it can be identically zero for $\beta > \beta_c$ only in the
thermodynamic
limit $V\to\infty$.
\item[ii)] Instead of $\langle \mu\,\bar\mu\rangle$ itself it proves convenient
to
use the quantity
\begin{equation}
\rho \mathop=\limits_{|d|\to \infty} \frac{1}{2}\ln
\langle \mu({\bf d},0)\,\bar\mu({\bf 0},0)\rangle
\label{eq3.6}\end{equation}
$\rho$ has less fluctuations than $\langle\mu\rangle$ itself, and is
independent
on the boundary conditions.
\item[iii)] If $\langle\mu\rangle$ tends to zero as a power as
$\beta\to\beta_c$
\begin{equation}
\langle\mu\rangle\mathop\simeq\limits_{\beta\to\beta_c} (\beta-\beta_c)^\delta
\label{eq3.7}\end{equation}
then, from the definition Eq.(\ref{eq3.6})
\begin{equation}
\rho \simeq \frac{\displaystyle \delta}{\displaystyle \beta-\beta_c}
\label{eq3.8}\end{equation}
\end{itemize}
Eq.(\ref{eq3.8}) can be translated in terms of correlation length $\xi$ and of
the
critical index $\nu$ by use of the relation
\begin{equation}
\xi^{-1} \simeq (\beta-\beta_c)^\nu
\label{eq3.9}\end{equation}
If $\xi \gg a$ ($a$ = lattice spacing) and $L \gg a$, then $\langle\mu\rangle$
is
approximatively independent of $a$ (finite size scaling):
\begin{equation}
\langle\mu\rangle \simeq L^{-\delta/\nu}\Phi(\frac{L}{\xi})
\label{eq3.10}\end{equation}
$\Phi$ is an analytic function at finite volume,
and Eq.(\ref{eq3.10})
tends to Eq.(\ref{eq3.7})
as $V\to\infty$. The Eq.(\ref{eq3.10}) implies
\begin{equation}
\rho L^{-1/\nu} = f((\beta-\beta_c) L^{1/\nu})
\label{eq3.11}\end{equation}
For lattices of different size $L$ the quantity $\rho L^{-1/\nu}$ must be a
universal function of the scaled variable $(\beta-\beta_c) L^{1/\nu}$. The
limit
$L\to\infty$ is thus extracted and the exponents $\delta$ and $\nu$ can be
determined. Typical data for $\rho(d)$ are shown in fig.1.
\par\noindent
\begin{minipage}{0.5\linewidth}
\epsfxsize0.85\linewidth
{\centerline{
\epsfbox{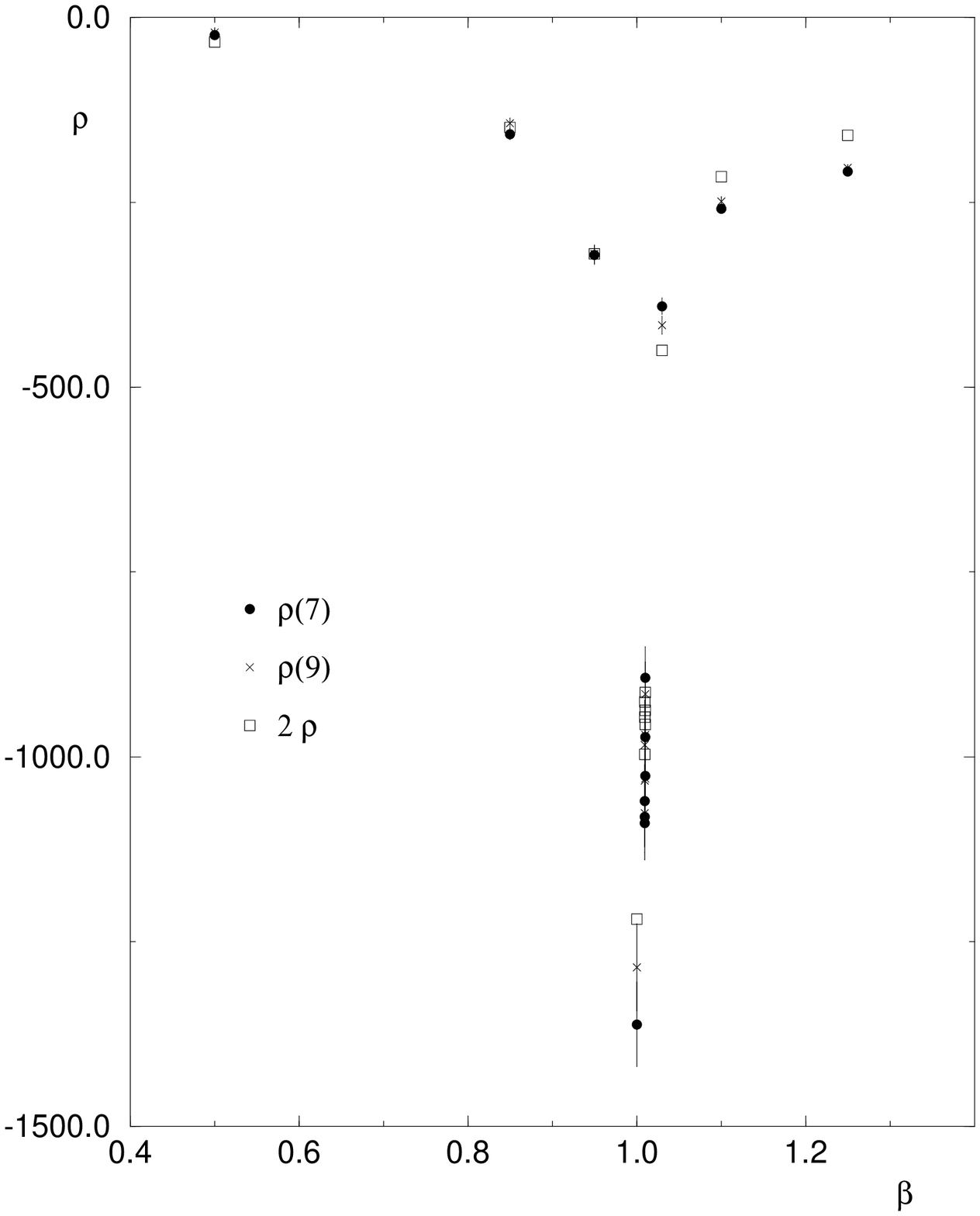}}}
\end{minipage}
\begin{minipage}{0.5\linewidth}
\epsfxsize0.85\linewidth
{\centerline{
\epsfbox{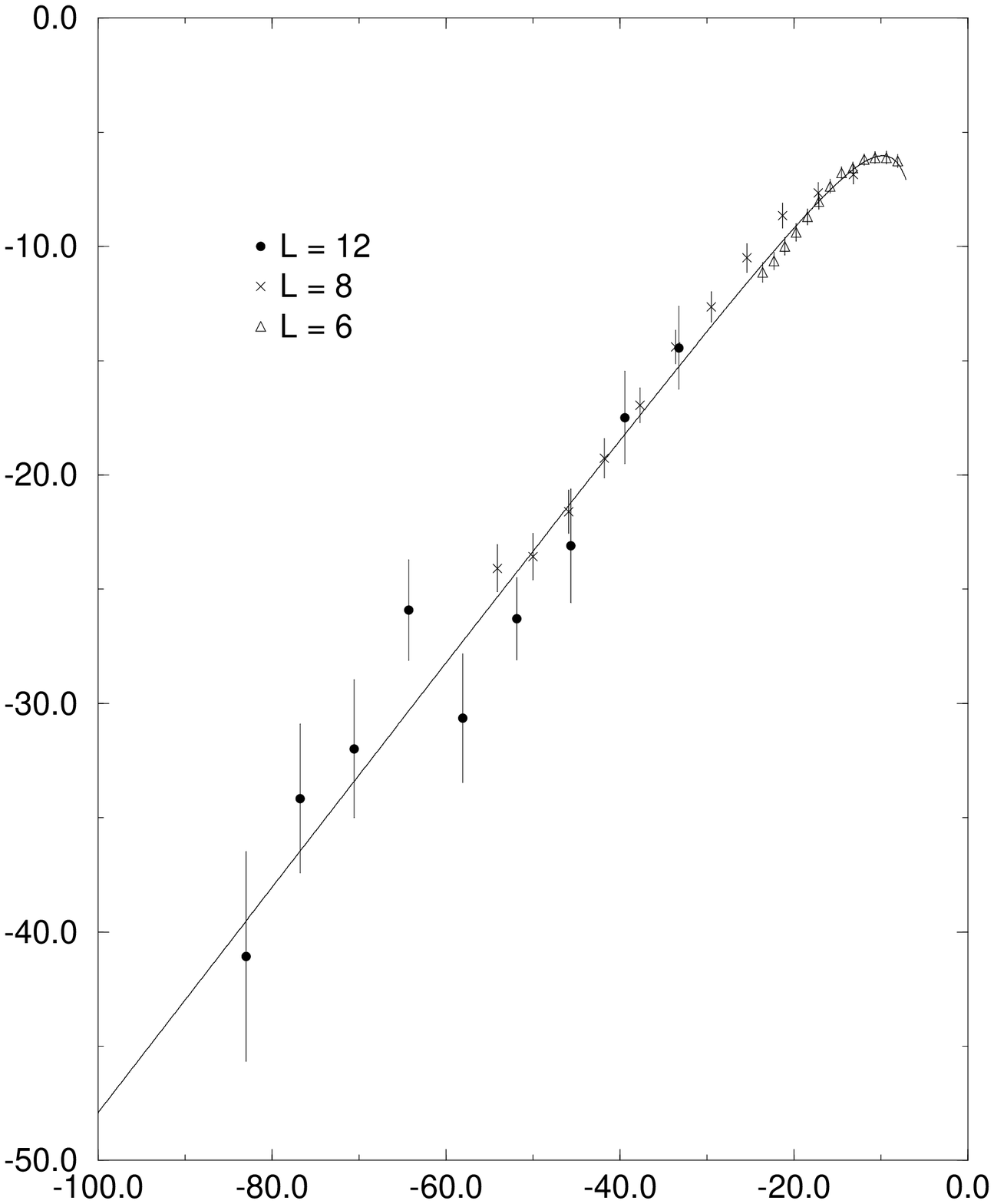}}}
\end{minipage}
\par\noindent
\begin{minipage}{0.5\linewidth}
{\centerline{Fig.1}}
\end{minipage}
\begin{minipage}{0.5\linewidth}
{\centerline{ Fig.2\quad $L^{\frac{1}{\nu}}/\rho$ vs
$(\beta-\beta_c) L^{\frac{1}{\nu}}$}}
\end{minipage}

The scaling
(Eq.(\ref{eq3.11})) is demonstrated in fig.2, where $L^{1/\nu}\rho^{-1}$ is
plotted
versus $L^{1/\nu}(\beta_c - \beta)$ for different lattice sizes. Data for
periodic
b.c. are well described by
\begin{equation}
\langle\mu\rangle \simeq L^{-\delta/\nu}
\left[\left( \beta_c - \beta) L^{1/\nu} + v_0\right)^2 +
v_1^2\right]^{\delta/2}
\label{eq3.12}\end{equation}
A best fit gives $\delta = 2.0\pm.2$
$\beta_c = 1.0111(1)$
$1/\nu = 3.97\pm.40$ $v_0\sim v_1 \sim 1$.
For $\beta < \beta_c$ vacuum is a dual superconductor.

\section{Dual superconductivity in $SU(2)$ gauge theory\cite{23}}
We have applied the construction described in sect.3 to detect dual
superconductivity in $SU(2)$ gauge theory. We have probed condensation of the
monopoles defined by two different abelian projections\cite{19}:
\begin{itemize}
\item[(a)] The abelian projection defined by diagonalizing as effective Higgs
field $\hat\Phi$ the Polyakov line.
\item[(b)] The abelian projection defined by diagonalizing a component (say
$F_{12}$) of the field strength $F_{0i}$.
\end{itemize}
For the projection (a) the relevant abelian field strength $F_{0i}$
(Eq.(\ref{eq2.1}))
is simply $F_{0i} = \hat\Phi^a G_{0i}^a$, since $D_0\hat\Phi =
0$. The operator $\mu$ [Eq.(\ref{eq3.1})] is constructed in terms of $\Pi_{i}$
$=$ $F_{0i}$ and the analysis of the $U(1)$ model is repeated.
A typical behaviour is shown in Fig.3, where $\rho$ is plotted vs $\beta$.

\par\noindent
\begin{minipage}{0.5\linewidth}
\epsfxsize0.85\linewidth
{\centerline{
\epsfbox{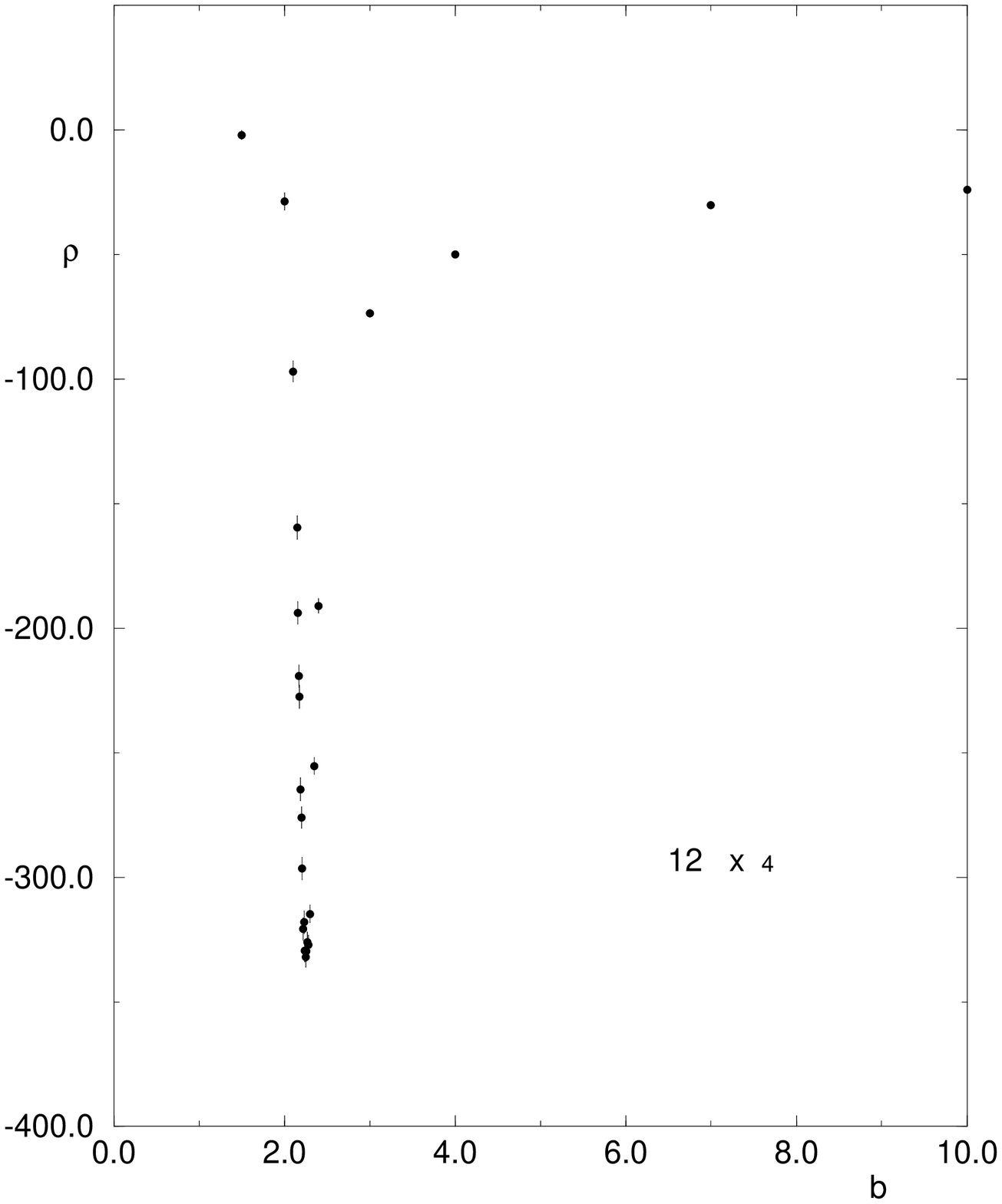}}}
\end{minipage}
\begin{minipage}{0.5\linewidth}
\epsfxsize0.85\linewidth
{\centerline{
\epsfbox{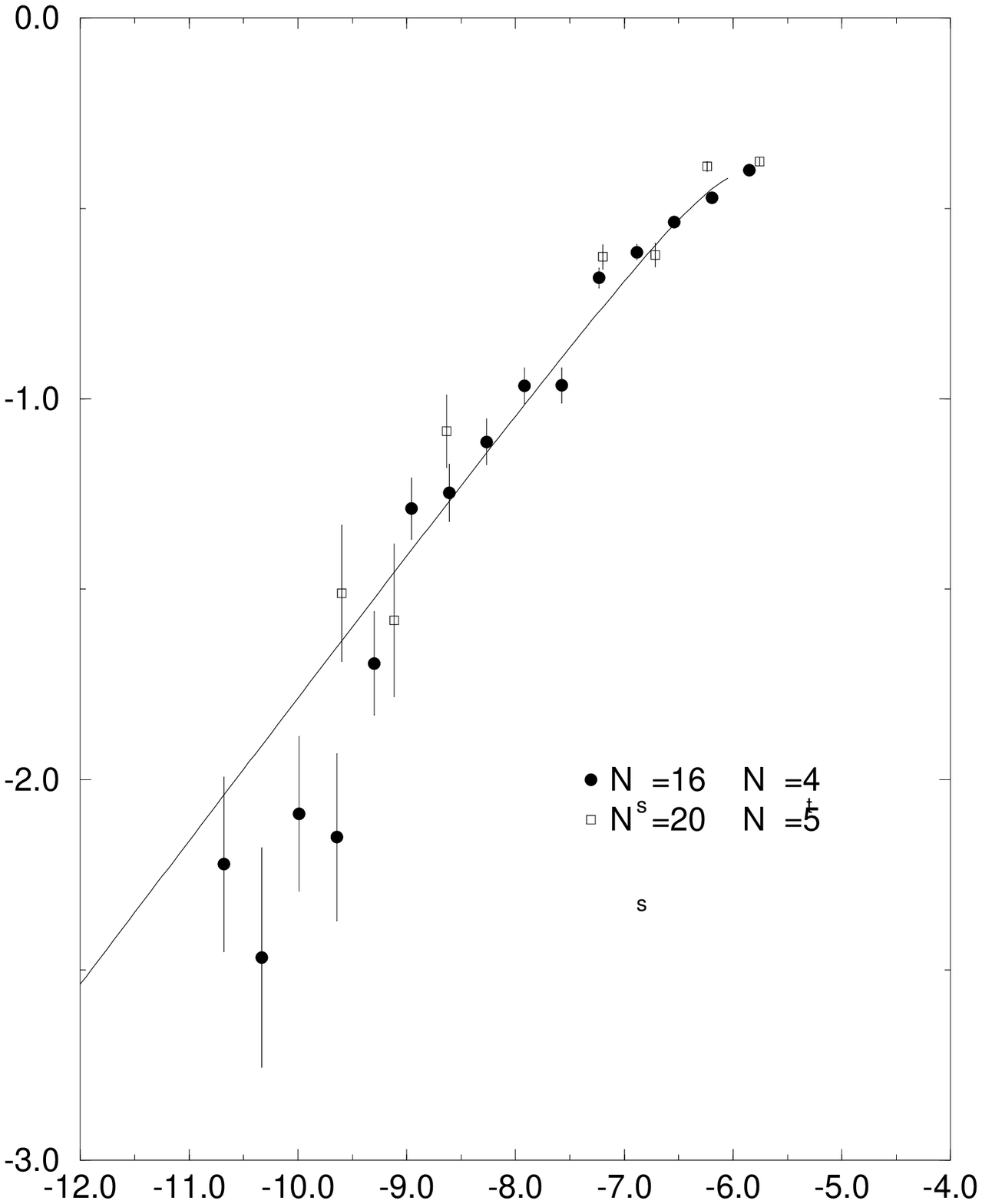}}}
\end{minipage}
\par\noindent
\begin{minipage}{0.5\linewidth}
{\centerline{Fig.3}}
\end{minipage}
\begin{minipage}{0.5\linewidth}
{\centerline{ Fig.4\quad $L^{\frac{1}{\nu}}/\rho$ vs
$(\beta/\beta_c-1) L^{\frac{1}{\nu}}$}}
\end{minipage}

The simulation is done on a asymmetric lattices $N_t = 4,6$,
$N_s = 16,20$. A clear signal is visible at the deconfining temperature. A
finite
size analysis confirms that condensation survives the limit $V\to\infty$
(fig.4).
The best fit gives:
\[
\nu \simeq 0.65\qquad \delta = 1.3\pm0.1\quad
\Delta\beta_c\equiv\beta_c(N_T = 6) -
\beta_c (N_T= 4) = 0.048\pm0.002\]
to be compared to $\Delta\beta_c = 0.07$ predicted by two loop
asymptotic scaling.
For the abelian projection (b) no signal is observed. There is no correlation
between the condensation of monopoles defined by this projection and
deconfinement.

\section{Concluding remarks}
\begin{itemize}
\item[(i)] We have demonstrated that the abelian projection which diagonalizes
the
Polyakov line defines monopoles condensing in QCD vacuum. The dual $U(1)$
corresponding to their charge is spontaneously broken and the QCD vacuum is a
dual
superconductor.
Recent observations that the abelian string tension in this projection is
almost
equal the usual string tension support our conclusion\cite{24a}.
\item[(ii)] Most of the work done in the literature on the role of monopoles in
confinement consists in correlating confinement to the density of monopoles or
of
monopoles world lines, as suggested by the pioneering work of ref.\cite{31} on
$U(1)$: a good review is contained in ref.\cite{20}. Of course the density of
monopoles is not a disorder parameter for dual superconductivity, in the sense
described in sect. 1, in the same way as the density of electrons or of Cooper
pairs is not for ordinary superconductors. In fact the density of monopoles,
contrary to $\mu$ (Eq.(\ref{eq3.4})), commutes with the monopole charge $M$,
and
cannot signal condensation.

\item[(iii)] Most of the work done in the literature has been done with the so
called ``maximal abelian'' projection\cite{27}. The monopoles defined by this
projection seem to be relevant to confinement, as evidenced also by the
detection
of persistent currents\cite{28,29}. The maximal abelian gauge presents less
lattice artifacts than others\cite{30}. We plan to investigate also this
projection by our method: a problem with computing power comes from the fact
that
the gauge is defined by a maximization which has to be repeated at each
updating
step in the computation of $\rho$.
\end{itemize}

In conclusion we have produced conclusive and direct evidence that
\begin{itemize}
\item[(i)] QCD vacuum is a dual superconductor.
\item[(ii)] not all the abelian projections are equally good to define the
monopoles relevant to confinement\cite{19}.
\end{itemize}

\end{document}